\documentclass[11pt]{article}

\usepackage{amsthm}
\usepackage{amsmath}
\usepackage{amssymb}
\usepackage[margin=1in]{geometry}
\usepackage[backref=page]{hyperref}
\usepackage{graphicx}
\usepackage{booktabs}
\usepackage{enumitem}
\usepackage{multirow}
\usepackage{fancyvrb}
\usepackage{listings}
\usepackage{float}

\hypersetup{
    colorlinks=true,     
    linkcolor=blue,      
    citecolor=blue,      
    filecolor=blue,      
    urlcolor=blue        
}

\newtheorem{remark}{Remark}

\newcommand{\cX}{\mathcal{X}}

\newcommand{\RR}{\mathbb{R}}
\newcommand{\CC}{\mathbb{C}}

\DeclareMathOperator{\Tr}{Tr}
\DeclareMathOperator{\conv}{conv}

\DeclareMathOperator{\id}{id}

\newcommand{\cvxpade}{\textsc{CvxQuad}}

\newcommand{\matlab}{\Verb}

\newcommand{\ket}[1]{|#1\rangle}

\newcommand{\braket}[2]{\langle#1|#2\rangle}
\newcommand{\proj}[1]{| #1\rangle\!\langle #1 |}

\newcommand{\Den}{\textsf{D}} 
\newcommand{\Lin}{\textsf{L}} 
\newcommand{\Sep}{\textsf{Sep}} 
\newcommand{\REE}{\textsf{REE}} 

\title{Efficient optimization of the quantum relative entropy}

\date{September 3, 2017}

\author{Hamza Fawzi\thanks{Department of Applied Mathematics and Theoretical Physics, University of Cambridge, UK.} \and Omar Fawzi\thanks{Laboratoire de l'Informatique du Parall\'elisme, ENS de Lyon, France}}

\begin{document}
\maketitle

\begin{abstract}
Many quantum information measures can be written as an optimization of the quantum relative entropy between sets of states. For example, the relative entropy of entanglement of a state is the minimum relative entropy to the set of separable states. The various capacities of quantum channels can also be written in this way. We propose a unified framework to numerically compute these quantities using off-the-shelf semidefinite programming solvers, exploiting the approximation method proposed in \textit{[Fawzi, Saunderson, Parrilo, Semidefinite approximations of the matrix logarithm, \texttt{arXiv:1705.00812}]}. As a notable application, this method allows us to provide numerical counterexamples for a proposed lower bound on the quantum conditional mutual information in terms of the relative entropy of recovery. 
\end{abstract}

\section{Introduction}

Many quantities in quantum information theory can be formulated as optimization problems involving the quantum relative entropy function:
\[
D(\rho\|\sigma) = \Tr[\rho(\log \rho - \log \sigma)].
\]
Examples of such problems include various notions of channel capacities, and relative entropy measures such as the \emph{relative entropy of entanglement} \cite{vedral1997quantifying} or the \emph{relative entropy of recovery} \cite{li2014squashed}.
Despite being a convex function, the relative entropy function $D$ is not natively supported by standard conic optimization solvers. To address this issue, the recent paper \cite{logapprox} proposed a new way to accurately approximate the relative entropy function via semidefinite programming.
This allows us to use off-the-shelf semidefinite programming solvers to solve convex optimization problems involving relative entropy.

The purpose of this paper is to survey some optimization problems in quantum information theory where the proposed method can be used. In particular we show how various quantum information theoretic quantities can be computed using semidefinite programming. The problems we consider in this paper include: 
\begin{itemize}
\item Capacity of classical-quantum channel
\item Entanglement-assisted classical capacity of quantum channel
\item Quantum capacity of degradable channels
\item Relative entropy of entanglement
\item Relative entropy of recovery
\end{itemize}
Some tailored algorithms have previously been developed for some of these quantities \cite{zinchenko2010numerical,sutter2014efficient} and we show that the SDP-based approximations of \cite{logapprox} are in general much faster in addition to being more flexible. As an application, we provide a numerical counterexample for an inequality that was recently proposed in \cite{li2014squashed,brandao2015quantum} concerning the relative entropy of recovery. Other applications of relative entropy optimization are discussed in \cite{chandrasekaran2013conic}.

The approximations of \cite{logapprox} were implemented in the Matlab package for CVX called \cvxpade~and is available at the URL:
\[
\text{\url{https://www.github.com/hfawzi/cvxquad/}}
\]
Table \ref{tbl:functions} shows the functions implemented in \cvxpade. For each problem we consider in this paper we provide the Matlab code using \cvxpade~that is needed to do the computation. Some of the code also requires the \texttt{quantinf} package \cite{cubittQuantinf} for density matrix manipulations.

\begin{table}[h!]
\centering
\begin{minipage}{\linewidth}
\begin{tabular}{llll}
\toprule
Function & Definition & & SDP blocks size\footnote{For each function, the constructed SDP consists of several blocks, each of size indicated in the column ($n$ is the dimension of the input $\rho$, and $\sigma$ if applicable). The number of such blocks is equal to $k+m$ where $k$ and $m$ are two parameters that control the accuracy of the approximations. The default setting in \cvxpade~is $k=m=3$ which we observed gives a good accuracy for most practical situations. We refer to \cite{logapprox} for more details on the role of $k$ and $m$.} \footnote{For the function \texttt{quantum\_rel\_entr}, the SDP consists more precisely of $m$ blocks of size $(n^2+1)\times (n^2+1)$ and $k$ blocks of size $2n^2\times 2n^2$. Also if either $\rho$ or $\sigma$ are constant, the SDP blocks have size $2n\times 2n$ instead.}\\
\midrule
\verb|quantum_entr| & $\rho\mapsto -\Tr[\rho \log \rho]$ & Concave & $2n\times 2n$\\
\verb|trace_logm| & $\rho\mapsto \Tr[\sigma \log \rho]$ ($\sigma \succeq 0$ fixed) & Concave & $2n\times 2n$\\
\verb|quantum_rel_entr| & $(\rho,\sigma) \mapsto \Tr[\rho(\log\rho - \log\sigma)]$ & Convex & $2n^2\times 2n^2$\\
\verb|quantum_cond_entr| & $\rho_{AB} \mapsto H(\rho_{AB}) - H(\Tr_A \rho_{AB})$ & Concave & $2n^2\times 2n^2$\\
\verb|op_rel_entr| & $(\rho,\sigma) \mapsto \rho^{1/2} \log\left(\rho^{1/2} \sigma^{-1} \rho^{1/2}\right) \rho^{1/2}$ & Operator convex & $2n\times 2n$\\
\bottomrule
\end{tabular}
\end{minipage}
\caption{List of functions supported by \cvxpade.}
\label{tbl:functions}
\end{table}

The main technical idea underlying \cite{logapprox} and \cvxpade~is a rational function approximation of the logarithm function which preserves operator concavity. Such a rational approximation is obtained by using quadrature on an integral representation of $\log$ (hence the name \cvxpade). We refer the reader to \cite{logapprox} for more information on how the approximation works.  Internally all the functions given in Table \ref{tbl:functions} are represented as semidefinite constraints and can thus be used with any off-the-shelf semidefinite solver.

The paper is organized as follows. In Section \ref{sec:notations} we introduce the notations that will be used in this paper. In Section \ref{sec:capacities} we consider the problem of computing various channel capacities. In Section \ref{sec:ree} we consider the computation of the relative entropy of entanglement. Finally, Section \ref{sec:rer} is concerned with the relative entropy of recovery where we show via numerical counterexamples that the relative entropy of recovery is not in general a lower bound on the quantum conditional mutual information, answering a question of~\cite{li2014squashed,brandao2015quantum}.

\section{Notations}
\label{sec:notations}

Given a finite-dimensional Hilbert space $A$ we denote by $\Den(A)$ the set of density operators on $A$, i.e., the set of Hermitian positive semidefinite matrices of size $\dim(A)$ with unit trace. Given $\rho \in \Den(A)$ we denote by $H(\rho) = -\Tr[\rho \log \rho]$ the von Neumann entropy of $\rho$. If $\sigma \in \Den(A)$ is another density operator on $A$ we let $D(\rho \| \sigma) = \Tr[\rho (\log \rho - \log \sigma)]$ be the quantum relative entropy between $\rho$ and $\sigma$. If $\rho$ is a bipartite state on $A\otimes B$, the conditional entropy $H(A|B)_{\rho}$ is defined by $H(A|B)_{\rho} := H(AB)_{\rho} - H(B)_{\rho}$.

We denote by $\Lin(A)$ the set of linear operators on $A$. If $A$ has dimension $n$ then $\Lin(A) \cong \CC^{n\times n}$. A quantum channel $\Phi$ with input space $A$ and output space $B$ is a linear map $\Phi:\Lin(A)\to \Lin(B)$ that is completely positive and trace-preserving.

\section{Capacities of quantum channels}
\label{sec:capacities}

\subsection{Classical to quantum channel}

Given a (finite) input alphabet $\cX$ and a finite-dimensional Hilbert space $A$, a classical-quantum channel is a map $\Phi : \cX \rightarrow \Den (A)$ which maps symbols $x \in \cX$ to density operators $\Phi(x)$. The capacity of such a channel is given by the solution of the following optimization problem \cite{holevo2012quantum}:
\begin{equation}
\label{eq:cq-capacity}
\begin{array}{ll}
\underset{p \in \RR^{\cX}}{\text{maximize}} & H\left(\sum\limits_{x \in \cX} p(x) \Phi(x) \right) - \sum\limits_{x \in \cX} p(x) H( \Phi(x) )\\
\text{subject to} & p\geq 0, \sum_{x \in \cX} p(x) = 1.
\end{array}
\end{equation}
where $H$ is the von Neumann entropy. The variable in \eqref{eq:cq-capacity} is a probability distribution $p$ on $\cX$.
One can formulate this optimization problem using the following CVX code, which uses the \texttt{quantum\_entr} function of \cvxpade~(for simplicity the code shown here is when $\cX$ consists of two elements):
\begin{lstlisting}[caption=Capacity of cq-channel,label=lst:cqcap]
% Input alphabet X={1,2}. 
% Generate rho1 and rho2 at random using quantinf's randRho function
rho1 = randRho(2);
rho2 = randRho(2);
cvx_begin
  variables p1 p2;
  maximize ((quantum_entr(p1*rho1 + p2*rho2) ...
               - p1*quantum_entr(rho1) - p2*quantum_entr(rho2))/log(2))
  p1+p2 == 1; p1 >= 0; p2 >= 0;
cvx_end
\end{lstlisting}
\paragraph{Accuracy} Since \cvxpade~is based on an approximation of the entropy function, we first check the accuracy of the method on a simple channel for which the capacity is known. Consider a channel on binary input alphabet $\cX=\{0,1\}$ where $\Phi(0)$ and $\Phi(1)$ are pure states 
\begin{equation}
\label{eq:cqpure} \Phi(x) = \proj{\psi_x} \quad \text{ where }  \quad \psi_x \in \CC^2, \;\; \braket{\psi_x}{\psi_x} = 1.
\end{equation}
In this this case it known that the capacity of the channel is $h_2((1+\epsilon)/2)$ where $h_2$ is the binary entropy function and $\epsilon = |\braket{\psi_0}{\psi_1}|$ (see \cite[Example 1]{holevo2012quantum}). We generate random channels by generating random states $\psi_0$ and $\psi_1$. Table \ref{tbl:cq_channel_1} shows the value of the capacity computed by solving \eqref{eq:cq-capacity} via \cvxpade, vs. the true value of the capacity. We see that the absolute error between the true value of capacity and the value obtained by running Listing \ref{lst:cqcap} is of the order of $10^{-6}$.

\begin{table}[ht]
\centering
\begin{tabular}{ccc}
\toprule
\textbf{True capacity} & \textbf{Computed value} & \textbf{Absolute error}\\
\midrule
0.775099 & 0.775100 & 1.47e-06\\
0.760555 & 0.760557 & 2.17e-06\\
0.713590 & 0.713593 & 2.29e-06\\
0.781838 & 0.781840 & 2.04e-06\\
0.487310 & 0.487312 & 1.98e-06\\
\bottomrule
\end{tabular}
\caption{Classical-to-quantum channel capacity of random channels of the form \eqref{eq:cqpure}. We see that the absolute error between the true value of capacity and the value obtained by running Listing \ref{lst:cqcap} is of the order of $10^{-6}$.}
\label{tbl:cq_channel_1}
\end{table}

\paragraph{Computation time} The paper \cite{sutter2014efficient} developed a tailored first-order optimization method to compute the solution of \eqref{eq:cq-capacity}. In Table \ref{tbl:cq_channel_2} we compare the running times of the approach of \cite{sutter2014efficient} and the SDP-based approach of \cvxpade. We see that the SDP-based approach scales better for channels with large output dimension.

\begin{table}[ht]
\centering
\begin{tabular}{llp{4cm}p{4cm}}
\toprule
\textbf{Input size} & \textbf{Output dim.} & \textbf{First-order method}\newline \textbf{of \cite{sutter2014efficient}} & \textbf{Approach based\newline on \cvxpade}\\
\midrule
 2 & 2 & 0.29s & 0.42s\\
 2 & 6 & 3.21s & 0.48s\\
 2 & 10 & 15.15s &  0.75s\\
 6 & 2 & 1.29s & 0.49s\\
 6 & 6 & 7.32s & 0.63s\\ 
 6 & 10 & 17.35s & 0.70s\\
 10 & 2 & 1.23s & 0.48s\\
 10 & 6 & 14.56s & 0.63s\\
 10 & 10 & 35.77s & 0.69s\\
 \bottomrule
 \end{tabular}
 \caption{Comparison of computation time for the first-order method of \cite{sutter2014efficient} and the SDP-based method of \cvxpade~for the capacity of classical-quantum channels. The number of iterations of \cite{sutter2014efficient} was set to achieve an \textit{a priori} accuracy of $10^{-2}$. The \textit{a posteriori} error ended up being around $10^{-4}$ for the different instances. The channel output dimension seems to have a strong effect on the method of \cite{sutter2014efficient}.}
 \label{tbl:cq_channel_2}
 \end{table}

\subsection{Entanglement-assisted classical capacity}

Consider a quantum channel $\Phi$ with input system $A$ and output system $B$. The \emph{entanglement-assisted classical capacity} denoted $C_{ea}$ quantifies the amount of classical bits one can transmit reliably through the channel, if the transmitter and receiver are allowed to share an arbitrary entangled state. The entanglement-assisted classical capacity of a quantum channel was shown in \cite{bennett2002entanglement} to have a simple maximization expression:
\begin{equation}
\label{eq:defCea}
C_{ea}(\Phi) = \max_{\rho \in \Den(A)} I(\rho,\Phi) 
\end{equation}
where $I(\rho,\Phi)$ is the so-called \emph{mutual information} of the channel $\Phi$ for the input $\rho$. This formula plays the same role as the well-known formula for the Shannon capacity of a classical channel. To define the mutual information $I(\rho,\Phi)$ let $U:A\rightarrow B\otimes E$ be a Stinespring isometry for $\Phi$ with environment $E$, i.e., such that $\Phi (X) = \Tr_{E}[U X U^*]$ for any operator $X$ on $A$. Then $I(\rho,\Phi)$ is defined as:
\begin{equation}
\label{eq:defmi}
I(\rho,\Phi) := H(B|E)_{U\rho U^*} + H(B)_{U\rho U^*}
\end{equation}
where $H(B|E)$ denotes the conditional entropy.

An important property of the mutual information defined in \eqref{eq:defmi} is that it is concave in $\rho$. This follows from concavity of conditional entropy which follows from convexity of relative entropy: indeed if $\sigma$ is a bipartite state on $BE$ then one can verify that
\begin{equation}
\label{eq:condentrconcave}
H(B|E)_{\sigma} = -D\left(\sigma_{BE} \| \sigma_{B} \otimes \id_{E} \right).
\end{equation}
Since $H(B|E)$ and $H(B)$ are concave in $\sigma = U\rho U^*$ it follows that $I(\rho,\Phi)$ is concave in $\rho$. See~\cite[Chapter 7]{holevobook} for more properties of $I(\rho,\Phi)$.

\begin{figure}[H]
  \centering
  \includegraphics[width=8cm]{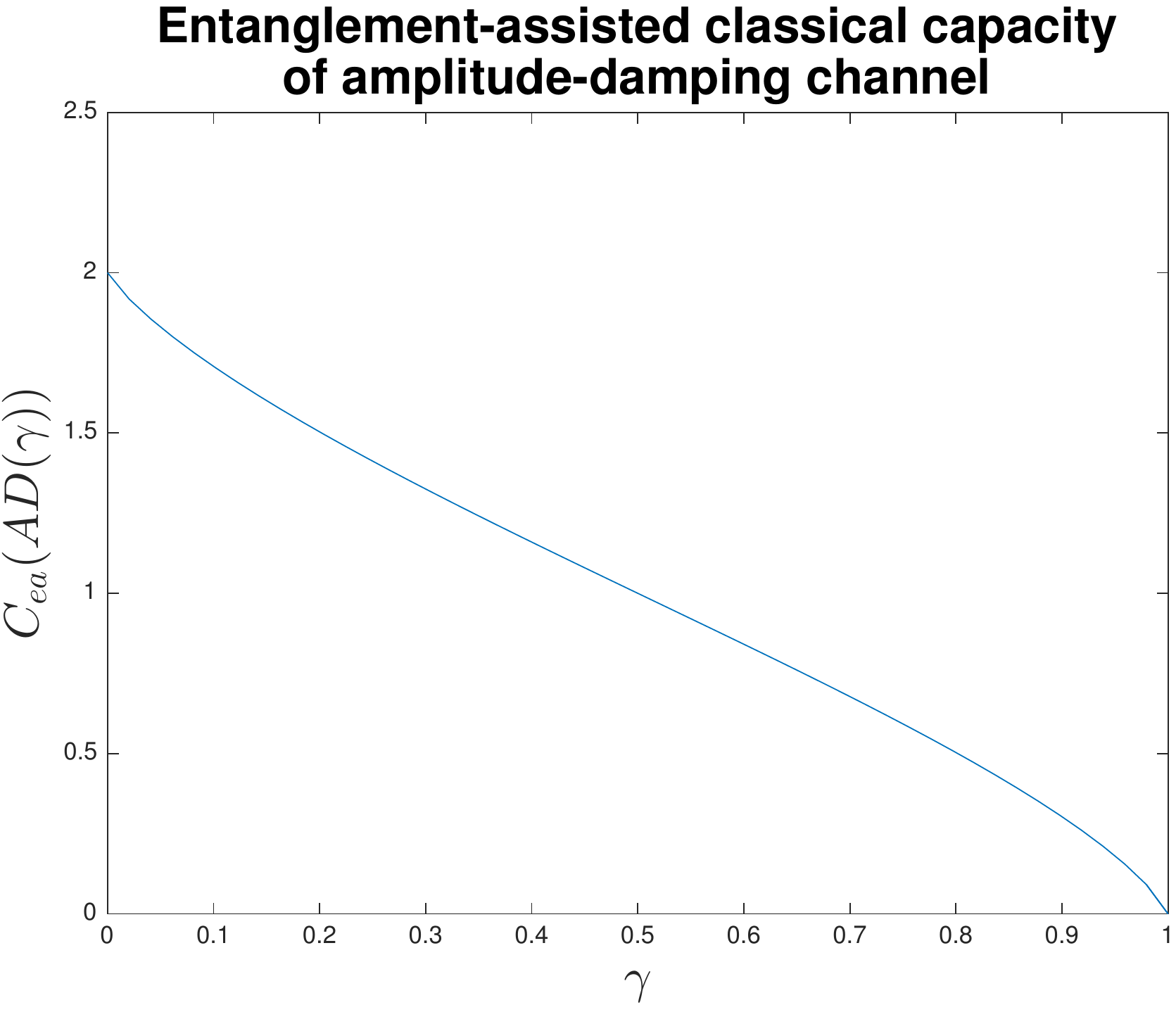}
  \caption{Entanglement-assisted classical capacity of amplitude damping channel}
  \label{fig:ea_capacity_AD}
\end{figure}

\paragraph{Illustration} Consider the so-called \emph{amplitude-damping} channel defined by (see \cite{giovannetti2005information} and \cite[Section 21.6.2]{wildebook}):
\begin{equation}
\label{eq:ADchannel}
\Phi(\rho) = A_0 \rho A_0^* + A_1 \rho A_1^*
\end{equation}
where
\[
A_0 = \begin{bmatrix} 1 & 0\\ 0 & \sqrt{1-\gamma} \end{bmatrix},
\quad
A_1 = \begin{bmatrix} 0 & \sqrt{\gamma}\\ 0 & 0\end{bmatrix}.
\]
The following Matlab/CVX script uses the \texttt{quantum\_cond\_entr} function of \cvxpade~to compute the entanglement-assisted capacity of the amplitude damping channel for the parameter $\gamma = 0.2$:
\begin{lstlisting}[caption=Entanglement-assisted classical capacity of a quantum channel,label=lst:eacap]
% Dimensions of input, output, and environment spaces of channel
na = 2; nb = 2; ne = 2;
% AD(gamma) = isometry representation of amplitude damping channel
AD = @(gamma) [1 0; 0 sqrt(gamma); 0 sqrt(1-gamma); 0 0];
U = AD(0.2);

cvx_begin sdp
  variable rho(na,na) hermitian;
  maximize ((quantum_cond_entr(U*rho*U',[nb ne]) + ...
               quantum_entr(TrX(U*rho*U',2,[nb ne])))/log(2));
  rho >= 0; trace(rho) == 1;
cvx_end
\end{lstlisting}
Figure \ref{fig:ea_capacity_AD} shows a plot of the entanglement-assisted classical capacity of the amplitude-damping channel as a function of $\gamma$.
Note that for this channel, \cite{giovannetti2005information} (see also \cite[Proposition 21.6.2]{wildebook}) gives a simplified expression of the entanglement-assisted classical capacity in terms of a univariate maximization problem.

\begin{remark}[Size of the semidefinite program to compute \eqref{eq:defCea}]
\label{rem:sizeCeaSDP}
As explained before, the function \texttt{quantum\_cond\_entr} will internally use a semidefinite programming approximation of the quantum conditional entropy function. The size of the SDP that is generated to express $I(\rho,\Phi)$ for a given channel $\Phi$ can be quite large. More precisely if we let $\dim(A),\dim(B),\dim(E)$ be respectively the dimensions of the input, output and environment spaces of the channel, then the SDP representation of the \texttt{quantum\_cond\_entr} term in Listing \ref{lst:eacap} (for the default approximation parameters) has a total of 6 positive semidefinite constraints, each of size $2(\dim(B)\dim(E))^2$ (more generally there are $k+m$ such constraints where $k$ and $m$ are the two parameters that control the accuracy of the approximation, see Table \ref{tbl:functions}, and \cite{logapprox} for details).

In the example of amplitude-damping channel, the environment has dimension $\dim(E) = 2$ and so the size of the each block is $2\times 4^2 = 32$. Some channels though have larger environment spaces, say $\dim(E) = 3$ (phase-damping channel) or $\dim(E) = 4$ (depolarizing channel) which may result in much larger SDPs. Many of the channels however have a lot of symmetry and this symmetry could be exploited to reduce the size of the resulting SDPs.
\end{remark}

\subsection{Quantum capacity of degradable channels}

The (unassisted) quantum capacity $Q(\Phi)$ of a quantum channel $\Phi$ is the number of quantum bits (qubits) one can reliably transmit over $\Phi$.
The quantum capacity of $\Phi$ admits the following expression \cite{holevo2012quantum}
\begin{equation}
\label{eq:QPhi}
Q(\Phi) = \lim_{n\rightarrow +\infty} \max_{\rho^{(n)}} \frac{1}{n} I_c(\rho^{(n)},\Phi^{\otimes n})
\end{equation}
where $I_c$ is the so-called \emph{coherent information} of a channel $\Phi$ for the input $\rho$:
\[
I_c(\rho,\Phi) := H(\Phi(\rho)) - H(\Phi^c(\rho)).
\]
In the expression above, $\Phi^c$ denotes a \emph{complementary channel} of $\Phi$ defined as follows: if $\Phi$ is a channel from $A$ to $B$ with environment $E$ and isometry representation $U:A\rightarrow B\otimes E$, then a complementary channel $\Phi^c : \Lin(A) \to \Lin(E)$ is given by:
\[
\Phi^c(\rho) = \Tr_{B}[U\rho U^*].
\]

In general, evaluating the quantum capacity \eqref{eq:QPhi} is difficult. However there is a class of so-called \emph{degradable} channels where the expression for $Q(\Phi)$ can be simplified \cite{devetak2005capacity}. A channel $\Phi$ is \emph{degradable} if there exists a channel $\Xi: \Lin(B) \rightarrow \Lin(E')$ with $E' \cong E$ such that $\Phi^c = \Xi \circ \Phi$. If a channel $\Phi$ is degradable one can show that the limit in \eqref{eq:QPhi} is not needed (i.e., one can fix $n=1$) and that $I_c(\rho,\Phi)$ is concave in $\rho$. The latter follows from concavity of quantum conditional entropy as we show now: assume $\Phi^c = \Xi \circ \Phi$ and let $W:B\rightarrow E' \otimes F$ be an isometry representation of $\Xi$. Then note that $\Phi^c(\rho) = \Xi(\Phi(\rho)) = \Tr_{F}[W \Phi(\rho) W^*]$. Thus we get that
\begin{equation}
\label{eq:Iccondentr}
\begin{aligned}
I_c(\rho,\Phi) &= H(\Phi(\rho)) - H(\Tr_{F}[W \Phi(\rho) W^*])\\
               &\overset{(a)}{=} H(W\Phi(\rho)W^*) - H(\Tr_{F}[W \Phi(\rho) W^*])\\
               &\overset{(b)}{=} H(F|E')_{W\Phi(\rho)W^*}
\end{aligned}
\end{equation}
where in $(a)$ we used the fact that $H(\sigma) = H(W\sigma W^*)$ since $W$ is an isometry and in $(b)$ we simply used the definition of conditional entropy. Since the conditional entropy is concave and $\rho\mapsto W\Phi(\rho) W^*$ is linear, it follows that $\rho\mapsto I_c(\rho,\Phi)$ is concave.

The previous derivation shows that for degradable channels, $I_c(\rho,\Phi)$ can be expressed in terms of conditional entropy and thus can be formulated using the functions in \cvxpade.

\begin{remark}
For general (non-degradable) channels, one can use the ``approximately degradable'' approach proposed in \cite{sutter2014approximate} to get an upper bound on the quantum capacity by solving a convex optimization problem similar to the one described above.
\end{remark}

\begin{figure}[H]
  \centering
  \includegraphics[width=7.5cm]{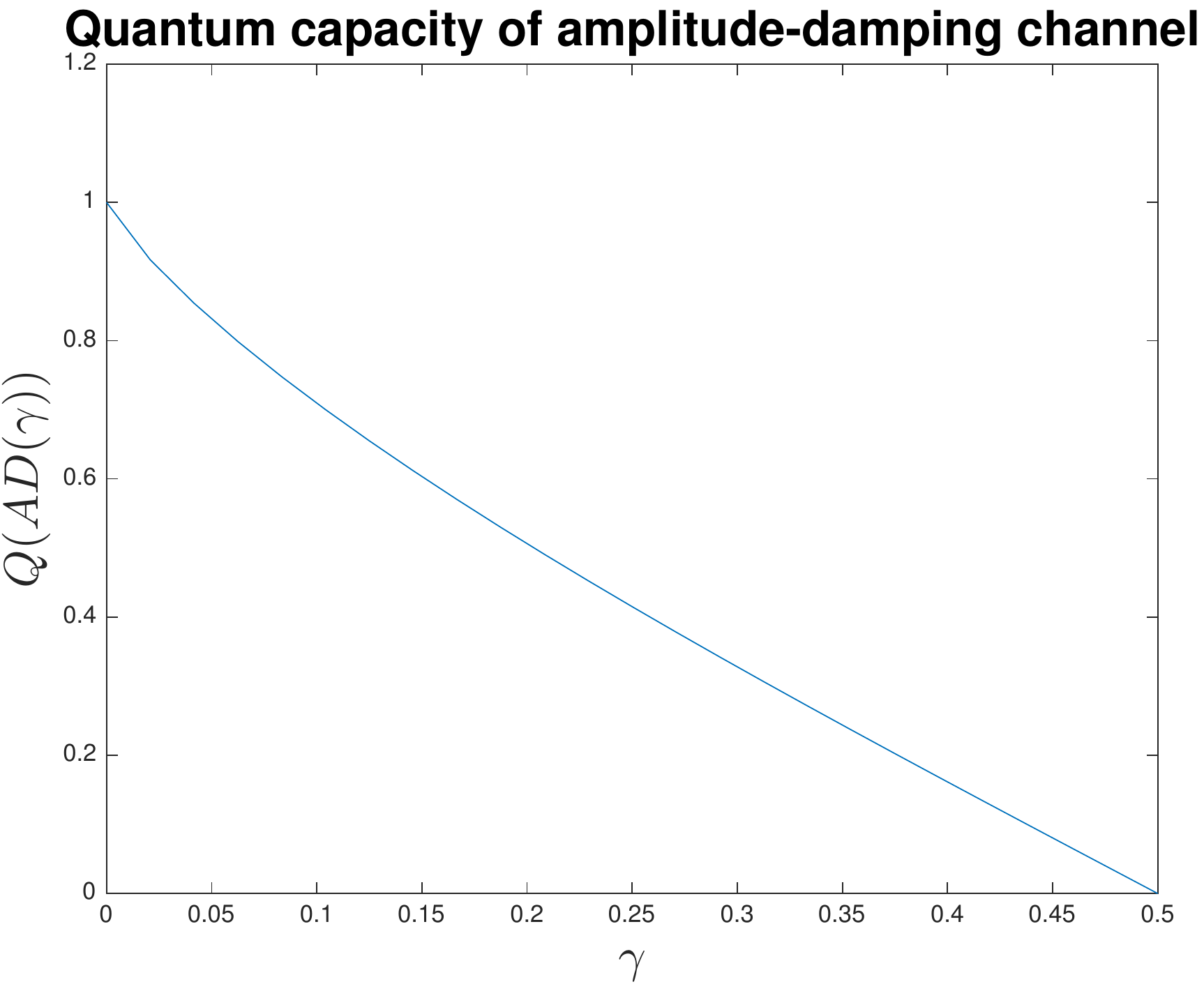}
  \caption{Quantum capacity of amplitude damping channel}
  \label{fig:q_capacity_AD}
\end{figure}

\paragraph{Illustration} As an illustration, we consider the amplitude-damping channel we saw in the previous section (see Equation \eqref{eq:ADchannel}). This channel is known to be degradable when $\gamma \leq 1/2$ and the degrading map in this case is an amplitude damping channel of parameter $(1-2\gamma)/(1-\gamma)$. Listing \ref{lst:qcapdeg} shows the Matlab/CVX script using the function \verb|quantum_cond_entr| of \cvxpade~to compute the quantum capacity of the amplitude-damping channel. Figure \ref{fig:q_capacity_AD} shows a plot of the quantum capacity of amplitude damping channels for $\gamma \in [0,1/2]$.

\begin{lstlisting}[float,floatplacement=H,caption=Quantum capacity of degradable channels,escapechar=!,label=lst:qcapdeg]
% Quantum capacity of degradable channels
% Example: amplitude damping channel
na = 2; nb = 2; ne = 2; nf = 2;
% AD(gamma) = isometry representation of amplitude damping channel
AD = @(gamma) [1 0; 0 sqrt(gamma); 0 sqrt(1-gamma); 0 0];
gamma = 0.2;
U = AD(gamma);
% Unitary representation of degrading map
W = AD((1-2*gamma)/(1-gamma));

% Ic(rho) = coherent information (see Eq. !\eqref{eq:Iccondentr}!)
Ic = @(rho) quantum_cond_entr( ...
                W*applychan(U,rho,'isom',[na nb])*W', [ne nf], 2)/log(2);

% Quantum capacity = maximum of Ic (for degradable channels)
cvx_begin sdp
  variable rho(na,na) hermitian
  maximize (Ic(rho));
  rho >= 0; trace(rho) == 1;
cvx_end
\end{lstlisting}

\section{Relative entropy of entanglement}
\label{sec:ree}

Let $\Sep$ denote the convex set of separable states on $A\otimes B$ i.e.,
\[
\Sep = \conv \left\{ \rho_A \otimes \rho_B : \rho_A \in \Den(A), \rho_B \in \Den(B) \right\}.
\]
Given a bipartite state $\rho$ in $\Den(A\otimes B)$ the \emph{relative entropy of entanglement} is defined as the distance from $\rho$ to $\Sep$, in the relative entropy sense \cite{vedral1997quantifying}:
\[
\REE(\rho) = \min_{\tau \in \Sep} D(\rho \| \tau).
\]
Since the set $\Sep$ is difficult to describe in general \cite{gurvits2003classical} we can get instead a lower bound on $\REE(\rho)$ by replacing the constraint that $\tau \in \Sep$, by the constraint that $\tau$ has a \emph{positive partial transpose}:
\begin{equation}
\label{eq:REE1}
\REE^{(1)}(\rho) = \min_{\tau \in \text{PPT}} D(\rho \| \tau).
\end{equation}
Similarly one can also define the quantity $\REE^{(k)}(\rho)$ where the separability constraint is replaced by the $k$'th level of the SDP hierarchy \cite{doherty2004complete}.

The value $\REE^{(1)}(\rho)$ can be computed using the CVX code shown in Listing \ref{lst:ree}, which uses the function \matlab|quantum_rel_entr| of the \cvxpade~package. 
\begin{lstlisting}[float,floatplacement=H,caption=Relative entropy of entanglement,label=lst:ree]
na = 2; nb = 2;
rho = randRho(na*nb);  % Generate a random bipartite state rho

cvx_begin sdp
  variable tau(na*nb,na*nb) hermitian;
  minimize (quantum_rel_entr(rho,tau)/log(2));
  tau >= 0; trace(tau) == 1;
  Tx(tau,2,[na nb]) >= 0;  % Positive partial transpose constraint
cvx_end
\end{lstlisting}
The paper \cite{zinchenko2010numerical} (see also \cite{girard2015erratum}) provides a numerical method to estimate $\REE^{(1)}(\rho)$ using a tailored cutting-plane method which constructs a polyhedral approximation of the function $\tau \mapsto \Tr[\rho \log \tau]$ by evaluating its derivatives at different points. 
Table \ref{tbl:ree_results} shows the result of computing $\REE^{(1)}(\rho)$ using the script above, for randomly generated $\rho$ of different sizes, as well as with the cutting-plane method of \cite{zinchenko2010numerical,girard2015erratum}. We see that the method based on \cvxpade~is much faster and scales better for states of large size.

\begin{table}[ht]
\centering
\begin{tabular}{lp{5cm}p{4cm}}
\toprule
$n = n_A \times n_B$ & \textbf{Cutting-plane approach}\newline \textbf{ of \cite{zinchenko2010numerical,girard2015erratum}} & \textbf{Approach based}\newline \textbf{ on \cvxpade}\\
\midrule
$4 = 2\times 2$ & 6.13 s & 0.55 s \\
$6 = 3\times 2$ & 12.30 s & 0.51 s \\
$8 = 4\times 2$ & 29.44 s & 0.69 s \\
$9 = 3\times 3$ & 37.56 s & 0.82 s \\
$12 = 4\times 3$ & 50.50 s & 1.74 s\\
$16 = 4\times 4$ & 100.70 s & 5.55 s\\
\bottomrule
\end{tabular}
\caption{Running times to compute the value $\REE^{(1)}(\rho)$ \eqref{eq:REE1} for random choices of bipartite states $\rho$ of size $n=n_A\times n_B$.
We see that on all the matrices tested, the \cvxpade~approach is much faster than the cutting-plane approach of \cite{zinchenko2010numerical,girard2015erratum}. Note that the algorithm \cite{zinchenko2010numerical,girard2015erratum} returns an interval of length $\epsilon$ guaranteed to contain the optimal value of \eqref{eq:REE1} (we set $\epsilon = 10^{-3}$ in the experiments). In all cases tested the value returned by the \cvxpade~code lied in that interval.}
\label{tbl:ree_results}
\end{table}

\section{Relative entropy of recovery and conditional mutual information}
\label{sec:rer}

Consider a tripartite state $\rho_{ABC}$ defined on the Hilbert space $A\otimes B \otimes C$. The conditional mutual information of $A$ and $C$ given $B$ is defined by
\[
I(A:C|B)_{\rho} = H(BC)_{\rho} + H(AB)_{\rho} - H(ABC)_{\rho} - H(B)_{\rho}.
\]
A state for which the conditional mutual information is zero is called a Markov chain $A-B-C$. In the case of a probability distribution, this corresponds to the random variables $A$ and $C$ being conditionally independent given $B$.
In the general quantum case, the word Markov chain is justified by the fact that such a state $\rho$ has the following property: starting with the marginal on $A \otimes B$, one can generate the system $C$ by only acting on the system $B$~\cite{petz88sufficiency}. In other words, it is possible to perfectly recover the $C$ system by acting only on $B$. 

In order to interpret the conditional mutual information operationally, it would be desirable to relate it to how well a state is recoverable~\cite{li2014squashed}. This leads us to consider states of the form
\begin{equation}
\label{eq:recoveredstates}
(\id_A \otimes \Lambda)(\rho_{AB}), \quad \Lambda: \Lin(B) \rightarrow \Lin(B\otimes C)
\end{equation}
where $\Lambda$ is a quantum channel. These correspond to tripartite states that are obtained by acting only on $B$. In~\cite{fawzi2015quantum}, it was shown that 
\begin{equation}
\label{eq:FRineq}
I(A:C|B) \geq \min_{\Lambda : B\rightarrow BC} \tilde{D}_{1/2}(\rho_{ABC} \| (\id_A \otimes \Lambda)(\rho_{AB})),
\end{equation}
where $\tilde{D}_{1/2}$ is a relative entropy measure of order $1/2$ that is smaller than the quantum relative entropy, but its exact definition is not important for this paper. 
 Inequality~\eqref{eq:FRineq} was then improved by~\cite{brandao2015quantum}, who replaced the relative entropy measure $\tilde{D}_{1/2}$ of order $1/2$ with the larger measured relative entropy $D^{\mathbb{M}}$. The authors of~\cite{li2014squashed} and~\cite{brandao2015quantum} naturally ask whether the inequality~\eqref{eq:FRineq} may be further improved to
\begin{equation}
\label{eq:BHOSineq}
I(A:C|B) \overset{?}{\geq} \min_{\Lambda : B\rightarrow BC} D(\rho_{ABC} \| (\id_A \otimes \Lambda)(\rho_{AB})).
\end{equation}

Since the right-hand side of~\eqref{eq:BHOSineq} --which we can call the \emph{relative entropy of recovery} of $\rho_{ABC}$-- is a quantum relative entropy optimization problem, we can test the inequality \eqref{eq:BHOSineq} numerically on any given state $\rho_{ABC}$. Listing \ref{lst:rerecov} shows the code to compute this quantity using the functions in \cvxpade. The following numerical examples show that the inequality \eqref{eq:BHOSineq} can be violated.

\begin{figure}[ht]
  \centering
  \includegraphics[width=9cm]{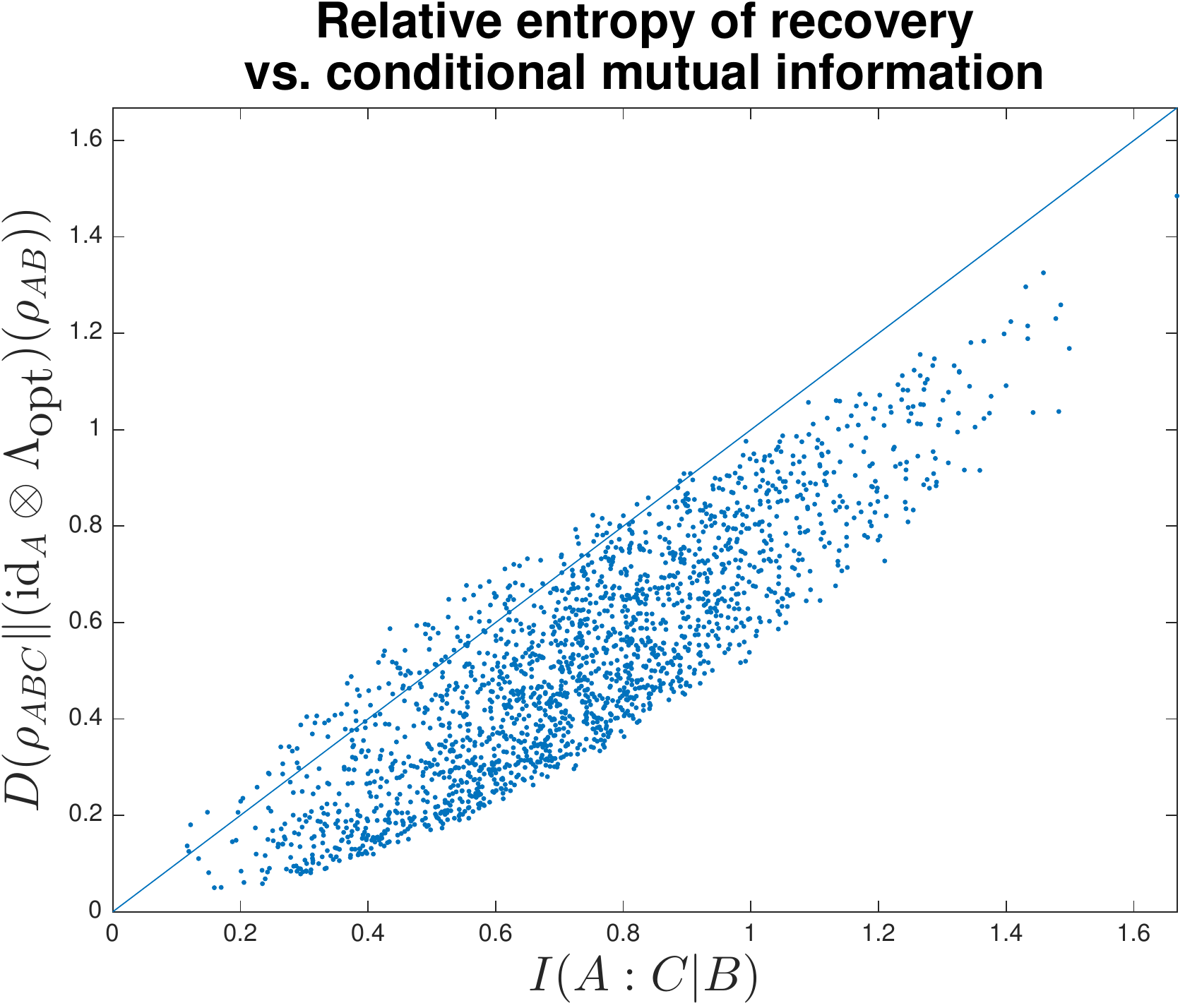}
  \caption{
  Plot of right-hand side of \eqref{eq:BHOSineq} vs. left-hand side for 2000 randomly generated rank-one (pure) states $\rho_{ABC}$. We see that some of the points violate the inequality.}
  \label{fig:ineq_BHOS_plots}
\end{figure}

\paragraph{Random pure states} We first tested \eqref{eq:BHOSineq} on random pure states where $\dim(A)=\dim(B)=\dim(C)=2$. Figure \ref{fig:ineq_BHOS_plots} shows a plot of the relative entropy of recovery vs. conditional mutual information for 2000 randomly generated states. We see that for some states the inequality \eqref{eq:BHOSineq} is violated.

\begin{figure}[ht]
  \centering
  \includegraphics[width=8cm]{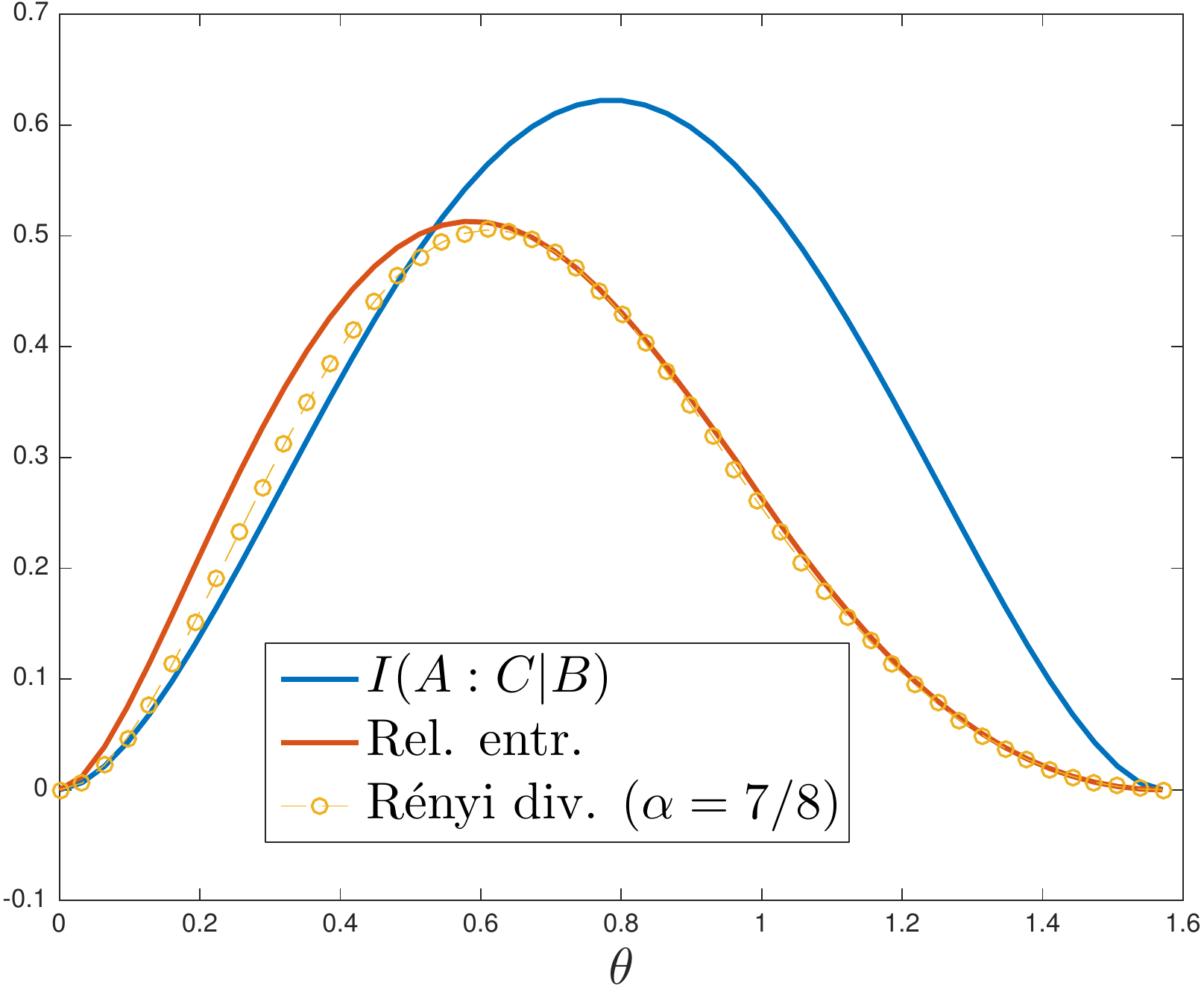}
  \caption{Condition mutual information and relative entropy of recovery for the state $\rho_{ABC}$ defined in Equations \eqref{eq:cex-rhoABC}-\eqref{eq:cex-psiABC}. For small values of $\theta$ the inequality \eqref{eq:BHOSineq} is violated.}
  \label{fig:plot_explicit_counterexample_ineq_BHOS}
\end{figure}

\paragraph{Explicit counter-example} We can also give an explicit state $\rho_{ABC}$ that violates \eqref{eq:BHOSineq}. Consider the following pure tripartite state:

\begin{equation}
\label{eq:cex-rhoABC}
\rho_{ABC} = \proj{\psi}_{ABC}
\end{equation}
where
\begin{equation}
\label{eq:cex-psiABC}
\ket{\psi}_{ABC} = \frac{1}{\sqrt{2}} \ket{0}_{B} \ket{00}_{AC}
                 + \frac{1}{\sqrt{2}} \ket{1}_{B}\left( \cos(\theta) \ket{01}_{AC} +\sin(\theta) \ket{10}_{AC} \right)
\end{equation}
Figure \ref{fig:plot_explicit_counterexample_ineq_BHOS} shows $I(A:C|B)$ and the relative entropy of recovery for the range $\theta \in [0,\pi/2]$. We have also included the quantum R{\'e}nyi divergence (for $\alpha = 7/8$) defined by
\[
\bar{D}_{\alpha}(\rho \| \sigma) = \frac{1}{\alpha-1} \log \Tr\left[\rho^{\alpha} \sigma^{1-\alpha}\right]
\]
which is a lower bound on the relative entropy $D$ and which can be optimized exactly using semidefinite programming for rational $\alpha$ \cite{fawzi2015lieb,sagnol2013semidefinite}. We see that there is a range of values of $\theta$ where the relative entropy of recovery is greater than or equal the conditional mutual information.

\begin{lstlisting}[float,floatplacement=H,caption=Relative entropy of recovery,label=lst:rerecov]
na = 2; nb = 2; nc = 2;
rhoABC = randRho(na*nb*nc);
rhoAB = TrX(rhoABC,3,[na nb nc]);

cvx_begin sdp

  % Channel \Lambda:B->BC (represented by its Choi-Jamiolkowski matrix)
  variable L_B_BC(nb^2*nc,nb^2*nc) hermitian
    
  % chanout_ABC will hold (id_A \otimes \Lambda)(rhoAB)
  variable chanout_ABC(na*nb*nc, na*nb*nc) hermitian
  
  % Objective function
  minimize (quantum_rel_entr(rhoABC, chanout_ABC) / log(2));

  % \Lambda must be trace-preserving and completely positive
  L_B_BC >= 0;
  TrX(L_B_BC,2,[nb nb*nc]) == eye(nb);
    
  % Form Choi matrix of id_A \otimes \Lambda given that of \Lambda
  % w is maximally entangled state on A
  w = zeros(na^2,1); w([1:na+1:na^2]) = 1; w = w*w';
  L_AB_ABC = sysexchange(tensor(w,L_B_BC),[2 3],[na na nb nb*nc]);
    
  % chanout_ABC is the result of applying id_A\otimes\Lambda to rhoAB
  chanout_ABC = applychan(L_AB_ABC,rhoAB,'choi2',[na*nb na*nb*nc]);

cvx_end
\end{lstlisting}

\paragraph{Acknowledgments} We would like to thank Tobias Sutter for providing us with the code of the method proposed in \cite{sutter2014efficient} to compute the capacity of classical-quantum channels.

\clearpage

\bibliography{sdp_rational_lieb}

\begin{thebibliography}{SSMER16}

\bibitem[BHOS15]{brandao2015quantum}
Fernando G. S.~L. Brand{\~a}o, Aram~W. Harrow, Jonathan Oppenheim, and Sergii
  Strelchuk.
\newblock Quantum conditional mutual information, reconstructed states, and
  state redistribution.
\newblock {\em Physical Review Letters}, 115:050501, Jul 2015.

\bibitem[BSST02]{bennett2002entanglement}
Charles~H. Bennett, Peter~W. Shor, John~A. Smolin, and Ashish~V. Thapliyal.
\newblock Entanglement-assisted capacity of a quantum channel and the reverse
  shannon theorem.
\newblock {\em IEEE Transactions on Information Theory}, 48(10):2637--2655,
  2002.

\bibitem[CS15]{chandrasekaran2013conic}
Venkat Chandrasekaran and Parikshit Shah.
\newblock Relative entropy optimization and its applications, 2015.
\newblock Available online at
  \url{http://users.cms.caltech.edu/~venkatc/cs_rep_preprint.pdf}.

\bibitem[Cub]{cubittQuantinf}
Toby Cubitt.
\newblock Quantinf package for {M}atlab.
\newblock \url{http://www.dr-qubit.org/Matlab_code.html}.

\bibitem[DPS04]{doherty2004complete}
Andrew~C. Doherty, Pablo~A. Parrilo, and Federico~M. Spedalieri.
\newblock Complete family of separability criteria.
\newblock {\em Physical Review A}, 69(2):022308, 2004.

\bibitem[DS05]{devetak2005capacity}
Igor Devetak and Peter~W. Shor.
\newblock The capacity of a quantum channel for simultaneous transmission of
  classical and quantum information.
\newblock {\em Communications in Mathematical Physics}, 256(2):287--303, 2005.

\bibitem[FR15]{fawzi2015quantum}
Omar Fawzi and Renato Renner.
\newblock Quantum conditional mutual information and approximate {M}arkov
  chains.
\newblock {\em Communications in Mathematical Physics}, 340(2):575--611, 2015.

\bibitem[FS17]{fawzi2015lieb}
Hamza Fawzi and James Saunderson.
\newblock Lieb's concavity theorem, matrix geometric means, and semidefinite
  optimization.
\newblock {\em Linear Algebra and its Applications}, 513:240--263, 2017.

\bibitem[FSP18]{logapprox}
Hamza Fawzi, James Saunderson, and Pablo~A. Parrilo.
\newblock Semidefinite approximations of the matrix logarithm.
\newblock {\em Foundations of Computational Mathematics}, 2018.
\newblock Package cvxquad at \url{https://github.com/hfawzi/cvxquad}.

\bibitem[GF05]{giovannetti2005information}
Vittorio Giovannetti and Rosario Fazio.
\newblock Information-capacity description of spin-chain correlations.
\newblock {\em Physical Review A}, 71(3):032314, 2005.

\bibitem[Gur03]{gurvits2003classical}
Leonid Gurvits.
\newblock Classical deterministic complexity of {E}dmonds' problem and quantum
  entanglement.
\newblock In {\em Proceedings of the Thirty-Fifth Annual ACM Symposium on
  Theory of Computing (STOC)}, pages 10--19. ACM, 2003.

\bibitem[GZFG15]{girard2015erratum}
Mark~W. Girard, Yuriy Zinchenko, Shmuel Friedland, and Gilad Gour.
\newblock Erratum: {N}umerical estimation of the relative entropy of
  entanglement [phys. rev. a 82, 052336 (2010)].
\newblock {\em Physical Review A}, 91(2):029901, 2015.

\bibitem[HG12]{holevo2012quantum}
Alexander~S. Holevo and Vittorio Giovannetti.
\newblock Quantum channels and their entropic characteristics.
\newblock {\em Reports on progress in physics}, 75(4):046001, 2012.

\bibitem[Hol13]{holevobook}
Alexander~S. Holevo.
\newblock {\em Quantum systems, channels, information: a mathematical
  introduction}, volume~16.
\newblock Walter de Gruyter, 2013.

\bibitem[LW14]{li2014squashed}
Ke~Li and Andreas Winter.
\newblock Squashed entanglement, k-extendibility, quantum {M}arkov chains, and
  recovery maps.
\newblock {\em arXiv preprint arXiv:1410.4184}, 2014.

\bibitem[Pet88]{petz88sufficiency}
D{\'e}nes Petz.
\newblock {Sufficiency of channels over von {N}eumann algebras}.
\newblock {\em Q. J. Math.}, 39(1):97--108, 1988.

\bibitem[Sag13]{sagnol2013semidefinite}
Guillaume Sagnol.
\newblock On the semidefinite representation of real functions applied to
  symmetric matrices.
\newblock {\em Linear Algebra and its Applications}, 439(10):2829--2843, 2013.

\bibitem[SSMER16]{sutter2014efficient}
David Sutter, Tobias Sutter, Peyman Mohajerin~Esfahani, and Renato Renner.
\newblock Efficient approximation of quantum channel capacities.
\newblock {\em IEEE Transactions on Information Theory}, 62(1):578--598, 2016.

\bibitem[SSWR14]{sutter2014approximate}
David Sutter, Volkher~B. Scholz, Andreas Winter, and Renato Renner.
\newblock Approximate degradable quantum channels.
\newblock {\em arXiv preprint arXiv:1412.0980}, 2014.

\bibitem[VPRK97]{vedral1997quantifying}
Vlatko Vedral, Martin~B. Plenio, Michael~A. Rippin, and Peter~L. Knight.
\newblock Quantifying entanglement.
\newblock {\em Physical Review Letters}, 78(12):2275, 1997.

\bibitem[Wil16]{wildebook}
Mark~M. Wilde.
\newblock From classical to quantum {S}hannon theory.
\newblock {\em http://arxiv.org/abs/1106.1445v7}, 2016.

\bibitem[ZFG10]{zinchenko2010numerical}
Yuriy Zinchenko, Shmuel Friedland, and Gilad Gour.
\newblock Numerical estimation of the relative entropy of entanglement.
\newblock {\em Physical Review A}, 82(5):052336, 2010.

\end{thebibliography}
\bibliographystyle{alpha}

\end{document}